\documentclass[preprint]{elsarticle}
\usepackage[utf8]{inputenc}
\usepackage{markdown}

\usepackage{braket}
\usepackage{amsmath}
\usepackage{amssymb}

\usepackage{graphicx}
\usepackage{dcolumn}
\usepackage{bm}

\begin{document}

\title{Ergodicity breaking in an incommensurate system observed by OTOCs and Loschmidt Echoes: From quantum diffusion to sub-diffusion.}

\author[1,2]{Fabricio S. Lozano-Negro}
\author[1,2]{Pablo R. Zangara}
\author[1,2]{Horacio M. Pastawski}

\address[1]{Facultad de Matem\'atica, Astronom\'{i}a, F\'{i}sica y Computaci\'on - Universidad Nacional de C\'ordoba, C\'ordoba, Argentina}
\address[2]{Instituto de F\'{i}sica Enrique Gaviola (CONICET-UNC), C\'ordoba, Argentina}

\begin{abstract} The metal-insulator transition (MIT), which includes Anderson localization and Mott insulators as extreme regimes, has received renewed interest as the many-body effects often constitute a limitation for the handling of quantum interference. This resulted in the field dubbed many-body localization (MBL), intensively studied theoretically and experimentally as understanding the appearance of equilibration and thermalization becomes relevant in dealing with finite systems. Here, we propose a new observable to study this transition in a spin chain under the ``disorder'' of a Harper-Hofstadter-Aubry-André on-site potential. This quantity, which we call \textit{zeroth-order gradient entanglement} (ZOGE) is extracted from the fundamental Fourier mode of a family of out-of-time-ordered correlators (OTOCs). These are just Loschmidt Echoes, where a field gradient is applied before the time reversal. In the absence of many-body interactions, the ZOGE coincides with the inverse participation ratio of a Fermionic wave function. By adding an Ising interaction to an XY Hamiltonian, one can explore the MBL phase diagram of the system. Close to the critical region, the excitation dynamics is consistent with a diffusion law. However, for weak disorder, quantum diffusion prevails while for strong disorder the excitation dynamics is sub-diffusive.
\end{abstract}

\begin{keyword}
Loschmidt Echoes \sep Many body localization \sep Incommensurate potential \sep Out of time order correlator \sep Quantum chaos \sep Anomalous spin diffusion
\end{keyword}

\maketitle

\section{\label{sec:level1} Introduction}

In the last decade, much effort has been made to study the many-body effects of quantum excitations in a lattice with a view to understand and control their dynamics in isolated quantum systems\cite{BIzSZ16}. In particular, while disorder may produce the absence of diffusion of excitations, \textit{i.e.} Anderson localization \cite{An58,An78}, weak many-body interactions usually tend to favor their spreading. Therefore, interactions should contribute to equilibration and thermalization. However, stronger interactions would lead to many-body localization (MBL) \cite{BAAl06b,GEi16,Im16}. In a clear limit situation, strongly interacting Fermions in a lattice yield a Mott insulator\cite{Mott68}. In these localized phases, ergodicity breaks down and local observables do not relax to thermal values\cite{R+I08}. Thus, MBL has been invoked as a mechanism that could prevent the excitation scrambling\cite{De91,RDO08,BAAl06b,Iy+Hu13}.

The first discussion on localization or metal-insulator transition (MIT) referred to many-spin systems \cite{An58,An78}. However, the complexity of the situation obliged P. W. Anderson to consider a single-particle propagating in a tight-binding lattice in presence of diagonal disorder. He showed that a disorder above some critical value would lead to absence of diffusion of the excitation, \textit{i.e.} a MIT. Two decades later, it became clear that in 1\(d\) and 2\(d\) systems, even the smallest disorder localizes all the single-particle eigenstates \cite{AAnLR79}. Thus, in order to have the full wealth of the localization phase transition with a finite critical disorder, one needs a 3\(d\) system. Even with these strong simplifications, handling the criticality of the phenomenon required extensive computational resources\cite{McKr93,RSc03}. 
However, a rich critical behavior is still possible in 1d for correlated disorder, as occurs in a periodic potential incommensurate with the lattice. 
A first physically relevant model was introduced by Harper and Hofstadter\cite{Har73,Hof76} to describe electrons under the effect of a magnetic field. This model, from now on HHAA, was generalized by Aubry and André \cite{AuAd80,WP84} who found that eigenstates are extended under a weak incommensurate potential but become localized at certain potential strength. This transition could be observed through the convergence of a perturbative expansion\cite{PWA83,WP84}, through the exponential decay of the Landauer conductance, \cite{McKr81,PaSW85} or by evaluating the inverse participation ratio (IPR) of the eigenstates\cite{Th77,Zg+Sa13,Xi+Da17,TS17,DoCa19,TS20}. Consequently, 1\(d\) incommensurate models have been frequently studied to mimic the Anderson transition of high dimensional disordered systems\cite{AuAd80}. 

The experimental and numerical characterization of the MBL transition in 3\(d\) has been extremely difficult to achieve. Hence, a natural shortcut is to use $1d$ incommensurate systems where there is already numerical and theoretical evidence that this transition survives the presence of interactions \cite{Iy+Hu13}. Particularly important experiments were performed in ultracold atoms arrays where periodic incommensurate potentials were implemented \cite{Sc+Bl15,Bo+Bl17}. Such systems were characterized, experimentally and theoretically, by computing the entanglement entropy\cite{Sc+Bl15,Xu+SwSa19}, the probability imbalances\cite{Sc+Bl15,Xu+SwSa19}, and dynamics of perturbed states \cite{YWWGC17,XuHZC20,B+RoT20}. More recently, the natural random disorder in certain $1d$ spin systems was studied through NMR \cite{WeiChCa18}. In this case, one can use implementations of the Loschmidt Echo (LE) techniques that involve the measurement of different observables following a perturbative pulse and a time reversed dynamics \cite{RfSaPa09}. Ref. \cite{WeiChCa18} used a combination of the LE procedures in disordered spin chains to evaluate different Out of Time Order Correlators (OTOCs). From these, they inferred the scrambling of collective spin excitations under different Hamiltonians and obtained an estimation of the entanglement entropy, which in turn, provides for an assessment of the MBL phase transition.

Since NMR can measure local correlation functions in the presence of time reversal \cite{ZMEr92,LUPa98}, we should be able to develop a new experimental strategy to monitor local excitations as they spread in a spin chain. This could become a unique tool to characterize the MBL transition in a 1\(d\) system. The idea is to label the degree of scrambling after the excitation has evolved under a Hamiltonian dynamics. This labeling is achieved by a Zeeman field gradient pulse that gives a different phase to each local component of the polarization. Thus, the effectiveness of the time reversal that follows this perturbation depends on each of these local phases. This concept has a loose analogy to the NMR’s Multiple Quantum Coherence sequences (MQC) \cite{MuPi86,DoMF00,MuPiM87}, where the Zeeman phase labels the component of the excitation along each of the spin projection sub-spaces. As in a MQC experiment, one can measure a set of OTOCs by applying different field gradients before time-reversal. This family of LEs can be Fourier transformed and its fundamental Fourier mode is our target magnitude: the \textit{zeroth-order gradient entanglement} (ZOGE). In absence of interactions, the ZOGE coincides with the \textit{inverse participation ratio} of the scrambled excitation. In the presence of interactions, the ZOGE roughly approximates the sum of the squares of the local components of the magnetization. 

In this paper, we develop the conceptual basis of the ZOGE procedure and test it numerically by studying the excitation dynamics in a spin chain with a ``disorder'' given by a Harper-Hofstadter-Aubry-André (HHAA) on-site potential. The spin dynamics is induced by the planar (or XY) Hamiltonian, which has been used to disclose the quantum nature of excitation dynamics in NMR experiment \cite{Ma+Er97,PLU95,PUL96}. Alternatively, full many-body dynamics is achieved by considering XXZ Hamiltonians, \textit{i.e.} the anisotropic spin-spin interactions. Thus, we designed a conceptually simple tool with clear experimental significance, which also has the advantage of being readily computed. This should allow us to explore the phase diagram of the system and to track the critical interactions with relatively low computational cost.

In section II we introduce the model and the details of our protocol. There, we also discuss the connection of the ZOGE observable with the inverse participation ratio of the excitation. 
In section III, we discuss how different initial conditions affect the MIT in our finite system. We also analyze the strategies to find a bound for the critical parameters in finite systems. 
In section IV we address the many-body case. We discuss the difference between the ZOGE and the sum of the squares of the local magnetization. On the basis of the computed dynamics we discuss how the power law decay of the survival probability of the excitation weakens as scattering processes become more relevant. Thus, we conclude that the asymptotic values of the ZOGE are the most robust quantities to detect localization. From these values we build a phase diagram that shows an increment of the critical disorder for small interactions that is consistent with experimental observation and previous theoretical hints. 
Section V has a final evaluation of our methods and results.

\section{The model} 

We consider the evolution of an initial excitation in a chain of \(N\) spins 1/2 arranged at distance \(a\). They interact under the Hamiltonian $\mathcal{H}=\mathcal{H}_{xy}+\mathcal{H}_{W}+\mathcal{H}_{I}$, where:
\begin{eqnarray}
\hat{\mathcal{H}}_{xy}&=&{J\sum_{n=1}^{N-1}}(\hat{S}_{n}^x \hat{S}_{n+1}^x+\hat{S}_{n}^y \hat{S}_{n+1}^y)\\
&=&{-\frac{J}{2}\sum_{n=1}^{N-1}}(\hat{S}_{n}^+ \hat{S}_{n+1}^-+\hat{S}_{n}^-\hat{S}_{n+1}^+),\\
\hat{\mathcal{H}}_{W}&=&-W\sum_{n=1}^{N} \cos{(2\pi qna+\phi)} \hat{S}_{n}^z,\\
\hat{\mathcal{H}}_{I}&=&U\sum_{n=1}^{N-1} \hat{S}_{n}^z \hat{S}_{n+1}^z\\
&=&U\sum_{n=1}^{N-1} (\hat{S}_{n}^+\hat{S}_{n}^-\hat{S}_{n+1}^+\hat{S}_{n+1}^--\dfrac{\hat{S}_n^z+\hat{S}_{n+1}^z}{2}+\dfrac{1}{4}),
\label{Hs}
\end{eqnarray}
The Hamiltonian $\mathcal{H}_{xy}$ contains only \(xy\) or in-plane interactions between the spins vectors. These produce the flip-flop processes. 
In order to better understand our model it is useful to transform it into a Fermionic representation using the prescription of Jordan and Wigner\cite{JW28}. Thus, the sum \(\mathcal{H}_{W}+\mathcal{H}_{xy}\) involves just on-site energies for a local Fermion density and single electron hopping amplitudes, both being just a product between a creation and an annihilation operator. Consequently, for $W=0$, it describes non-interacting spinless Fermions, which in an infinite chain would have the dispersion relation
\begin{equation}
  \varepsilon^o_{k_x}=-J\cos{[k_xa]},
\end{equation}
where \(a\) is the lattice constant and $k_x$ the wave vector along \(x\). The effect of pseudo-random ``disorder'' is introduced by $\mathcal{H}_{W}$, which here is an incommensurate on-site potential in a 1-d chain that extends along the \(x\)-direction. This Hamiltonian, with \(W=J\), was introduced by Harper and Hofstadter\cite{Har73,Hof76} to describe each of the $k_y$ electronic eigenstates of an infinite lattice laying in the \(xy\)-plane with a magnetic field in the direction $z$. Such field yields a flux \(\Phi=Ba^2\) per ``plaquette" which must be compared with the quantum flux $\Phi_0=2\pi\hbar c/e$ yielding the ratio of $q=\Phi/\Phi_0$. The wave vector becomes a phase that determines the center \(x_0\) of a semi-classical orbit that confines the electron along \(x\) as \(x_0/a=k_ya/q=\phi/q\). Clearly, the Fermion mass becomes defined at the long wavelength limit \(2a^2J/\hbar^2\rightarrow 1/m\) and the cosine potential 
\[-W\cos{(2\pi qna+\phi)}\rightarrow (1/2)m \omega_H^2(x-x_0)^2\]
 becomes the harmonic potential with $\omega_H$ the cyclotron frequency that yields the Landau levels. In the lattice one gets a fractal energy spectrum with the characteristic form of a Hofstadter's Butterfly\cite{Hb76}. If the system is an infinite strip along \(y\) and has a finite length \((N+1)a\) along \(x\), localized states appear at the extremes of the chain that are identified with the edge states of the continuous model \cite{RTJH83}. 

A few years before, Aubry and André had considered families of periodic potentials in a one dimensional infinite chain with a variable strength \(W\). They realized that if the period of the on-site potential is \textit{incommensurate} with \(a\), \textit{i.e.} if \(q\) is not a rational number, the Hamiltonian shows a self-dual symmetry between Bloch states along \(x\) and the localized states, \textit{i.e.} the Hamiltonian has the same form when it is represented using a local basis than in the momentum space\cite{AuAd80,WP84}. This implies that \textit{all} the eigenstates are extended, \textit{i.e.} local in the momentum space, for $W/J<1$ and that are localized, \textit{i.e.} extended in the momentum space, for $J/W<1$. Thus, a phase transition occurs at $W_c=J$. This justifies interpreting the pseudo-random sequence generated by $\cos{(2\pi qn+\phi)}$ as a \textit{model for a disordered potential}. In our case we choose one of the quadratic irrationals, $q=(1+\sqrt{5})/2=1.618...$, the golden ratio, whose irrational period can be progressively approximated by the Fibonacci sequence ratios {2/3, 3/5, 5/8, 8/13, ...} where the denominators are the corresponding periods commensurate with the lattice. This \(q\) turns to be \textit{the most irrational number}\cite{Ph15}, in that it is the least well approximated by rationals, as can be seen by truncating its continued fraction expansion \(q=1/(1+1/(1+1/(1+...)))\). Since we use finite chains, we choose $\phi$ as an arbitrary phase that produces different ``disorder'' realizations. 

The Ising Hamiltonian $\mathcal{H}_{I}$ is a many-body interaction since in the Fermionic representation it requires four Fermion operators. Clearly, this term, by itself, is not able to induce dynamics. Thus, when its strength dominates, any excitation remains essentially frozen in a sort of spin-glass phase. Nevertheless, when the interactions are added to the otherwise localized excitations, the term $\mathcal{H}_{I}$ provides multiple pathways through the Hilbert space allowing it to continue its quantum diffusion process \cite{Zg+Pa13,ZgLPa15}. 

In what follows, we propose a new experimental method to quantify the Aubry-André transition that remains valid when the interaction term is turned on. It is based on the concept of non-local out-of-time order correlators (OTOCs) that labels through perturbation an already scrambled excitation before imposing a time reversal procedure. In contrast to recent works on HHAA model \cite{YWWGC17,XuHZC20,B+RoT20} where the LEs have been used to study the dynamics of an eigenstate after a quench in the Hamiltonian, here we return to the original experimental LE. It is loosely inspired by the Multiple Quantum Coherence (MQC) pulse sequence used in NMR \cite{DoMF00} as we describe below.

In order to fix ideas we consider the idealized ingredients of an NMR experimental set-up \cite{ZgDLPa12}, as other experimental set ups are often adaptions of these concepts. One assumes a thermal energy \(k_BT=1/\beta\) which is very high as compared with the Zeeman frequency, \(\omega_o\), and the relevant parameters in the Hamiltonian, \(k_BT\gg \omega_o\gg J,U\). As in the polarization echo experiment, a local probe, e.g. a rare nucleus, is connected to individual spin, say site \(0-\rm{th}\). \cite{LUPa98} One may ensure that the polarization of each other spin cancels out. Therefore, the initial state is represented by the density matrix:
\begin{eqnarray}
\rho_0 = \frac{(\hat{\mathbb{I}}+\beta\omega_o \hat{S}^{z}_0)}{\rm{Tr}[{\mathbb{I}+\beta\omega_o \hat{S}^{z}_0}]}.
\end{eqnarray}
Since $\mathbb{I}$ does not contribute to the dynamics, $\rho_0 \propto S^z_0$ describes a pseudo-pure state \cite{CoFH97}. In an actual NMR experiment one uses a macroscopic sample where multiple distant sites are addressed simultaneously. However, this only yields a more robust observable that average out unwanted quantum fluctuations that characterize individual measurements.

The nature of the excitation is better understood in terms of the spin raising and lowering operators that are readily mapped into Fermionic creation and annihilation operators. These operate on the thermal equilibrium \(N\)-spin state described by $|\Psi_{\mathrm{eq}}\rangle$\cite{ZgDLPa12}. Thus, 
\begin{eqnarray}
  |\Psi_{0}\rangle&=&\frac{\hat{S}_{0}^{+}|\Psi_{\mathrm{eq}}\rangle}{\left\vert
 \langle\Psi_{\mathrm{eq}}|\left. \hat{S}_{0}^{-}\hat{S}_{0}^{+}\right\vert
 \Psi_{\mathrm{eq}}\rangle\right\vert ^{1/2}}\\
 &=&\sum_{r=1}^{2^{N-1}}
\frac{e^{\mathrm{i}\phi_{r}}}{2^{(N-1)/2}}\left\vert
\uparrow_{0}\right\rangle \otimes\left\vert \beta_{r}\right\rangle,
\end{eqnarray}
The denominators ensure the proper normalization of the state, $\phi_{r}$ is a random phase and $\left\vert \beta_{r}\right\rangle$ describes states of the form:
\begin{eqnarray*}
  \left\vert \beta_{r}\right\rangle &=&\left\vert s_{1}\right\rangle
\otimes\left\vert s_{2}\right\rangle \otimes\left\vert
s_{3}\right\rangle \otimes...\otimes\left\vert s_{N-1}\right\rangle \;\;\;\\ \\&\mathrm{with}& \;\;\; \left\vert s_{k}\right\rangle \in\left\{ \left\vert
\uparrow\right\rangle ,\left\vert \downarrow\right\rangle \right\} \,.
\end{eqnarray*}
This description assumes that in thermal equilibrium all correlations have already decayed and hence the phases can be considered random numbers. Of course, observed values will be subject to quantum fluctuations, as occur in an actual individual experiment. However, this noise does not survive ensemble average. 
Particularly useful is the excited subspace with the maximally negative spin projection, where states of the computational basis have the form
\begin{eqnarray}
	\vert \beta_{n}\rangle &=& \vert \downarrow_{0}\rangle
\otimes\left\vert \downarrow_{1}\right\rangle \otimes...\left\vert
\uparrow_{n}\right\rangle ...\otimes\left\vert \downarrow_{N-1}\right\rangle,\\
	\hat{\mathcal{H}}|\alpha_{k} \rangle&=& \varepsilon_{k}|\alpha_{k} \rangle=\varepsilon_{k}\sum_{n=1}^{N}a_{kn}|\beta_{n}\rangle.
  \end{eqnarray}
It is the only sub-space where the single particle dynamics controlled by the energies \(\varepsilon_k\) persists even in the case \(U\neq 0\). Thus, the excitation dynamics is described by the correlation function
\begin{eqnarray}
c_{n\vert 0}(t) &=&\langle  \beta_{n} \vert \exp[-\textrm{i}\hat H t] \left\vert \beta_{0}\right\rangle\\
&=&\sum_{k=1}^N \exp[-\textrm{i}\varepsilon_k t]a_{k n }^*a^{\  }_{k0}
\end{eqnarray}
in terms of the single-particle eigen-energies \( \varepsilon_k\).

In the general case, the excess polarization evolves for a time $t$. Then, every local component of the state is perturbed by the action of a pulse of field gradient $\hat{\mathcal{H}}_{g}=\sum_n n \varphi\hat{S}_{n}^{z}$. Here, the phase $\varphi$ is the product of the pulse lapse and the strength of the field gradient. The effect is readily understood by considering the maximally negative spin projection space of dimension $\binom{N}{1}$ discussed above, and the minimal spin projection subspace of spin projection 1/2 and dimension $\binom{N}{(N-1)/2}$. In order to interpret this effect, we resort to the non-interacting case where the excess local polarization is identified with a single particle local probability \cite{PLU95}. This would imply that this perturbation labels each local component of the total polarization with a phase $n\varphi$. After this, the system evolves backwards in time (\textit{i.e.} it evolves under $-\mathcal{H}$ during a further time $t$). One then registers the amount of magnetization comes back to the initial spin. Thus, identifying the Heisenberg operator 
\begin{equation}
\hat{\Phi}_\varphi(t)=e^{\textrm{i}t\hat{\mathcal{H}}}e^{-\textrm{i} \varphi \hat{\mathcal{H}}_{g}}e^{\textrm{-i}t\hat{\mathcal{H}}},
\end{equation}
where, from now on, we consider $\hbar=1$.

The Loschmidt echo (LE) under a sudden perturbative pulse has the clear form of an Out of Time Order Correlator,
\begin{eqnarray}\label{eq:LE}
 M(t,\varphi)  &=& \frac{\langle \Psi_{\rm{eq}}| \hat{\Phi}_{\varphi}^{\dagger}(t )\hat{S}_{0}^{-}\hat{\Phi}_{\varphi}(t)\hat{S}^{+}_0|\Psi_{\rm{eq}}\rangle}{\left\vert
  \langle\Psi_{\mathrm{eq}}|\left.  \hat{S}_{0}^{-}\hat{S}_{0}^{+}\right\vert
  \Psi_{\mathrm{eq}}\rangle\right\vert}\\
 &\equiv&\langle \hat{\Phi}_{\varphi}^{\dagger}(t)\hat{S}^{z}_0 \hat{\Phi}_{\varphi}(t)\hat{S}^{z}_0\rangle_\beta.
\end{eqnarray}
We used two alternative notations to stress the physical meaning of these mathematical objects which remains somewhat obscure in the literature. The first line has a clear interpretation in terms of the polarization dynamics initiated by a spin raising operator on a given thermal state, $|\Psi_{\mathrm{eq}}\rangle$ without net polarization which becomes \textbf{excited} $|\Psi_{\mathrm{0}}\rangle=\hat{S}^+_0|\Psi_{\mathrm{eq}}\rangle$) at the initial time. Then, it \textbf{evolves forward} in time under a Hamiltonian dynamics $e^{(-\textrm{i} t\hat{\mathcal{H}})}$ that scrambles the excitation. The system is \textbf{perturbed} by a gradient field pulse $e^{\textrm{-i}\varphi \hat{\mathcal{H}}_{g}}$ and then after a sudden change in the Hamiltonian sign it \textbf{evolves backwards} in time as $e^{\textrm{i} t\hat{\mathcal{H}}}$. As the perturbation prevents a perfect return, its effect is \textbf{detected} as the failure in reaching the original polarization. The second line follows the most standard notation in spin dynamics were $\langle\cdot\rangle_\beta$ means properly normalized ensemble average ($\textrm{Tr}[\cdot]$), the equivalence between these two views in discussed in \cite{ADLP08}.

The density matrix representation allows us to specify the relationship of this LE with an OTO commutator, 
\begin{equation}
\langle [\hat{S}(t),\hat{S}^z_0(0)]^\dagger [\hat{S}(t),\hat{S}^z_0(0)] \rangle_{\beta} = 2\langle \hat{S}^z_0(0){}^2\rangle_{\beta} -2 M(t,\varphi).
\end{equation}
In our case, the degree of scrambling of $S_0^z(t)$ within each total spin projection subspace is labeled by an instantaneous spin rotation $\Phi =e^{-\rm{i} \varphi \hat{\mathcal{H}}_{g}}$ around a Zeeman field gradient. This makes a crucial difference with the MQC sequence where a uniform rotation would have no observable effect when applied after a dynamics that conserves total spin projection.

As in MQC we will use the Fourier transform of the quantity $M(t,\varphi)$ with respect to the variable $\varphi$. For each evolution time, this yields the amplitudes $\tilde Q_n$, which we call \textit{gradient entanglement amplitudes} (GEA),
\begin{equation}
\tilde Q_n(t)=\frac{1}{2\pi} \int_{0}^{2\pi} M(t,\varphi) e^{-\rm{i}n\varphi} d\varphi.
\end{equation}
These positive coefficients encode information on the degree of the excitation. This builds up as the entanglement of the excitation's different local components at a distance \(na\). Thus, one expects that in a localized regime only low indexes \(\tilde Q_n\)'s are important. This contrasts with the delocalized regime where the $\tilde Q_n$ have a broader support that indicates the spreading of the magnetization. This is particularly clear in absence of interactions ($U=0$). Then, the value at the center of the distribution, $\tilde Q_0$, is equal to an inverse participation ratio (IPR) of the dynamical state, which corresponds to the sum of the squared local magnetizations. That is,
\begin{eqnarray}
\tilde Q_{0}\underset{U\rightarrow 0}{\longrightarrow } Q_0(t)&=&\sum_{n} \langle \hat S_n^z(t)\rangle_\beta^2 =\sum_n {S^z_n}^2\\
&=&\sum_{n} |c_{n \vert 0}(t)|^4=\rm{IPR_t}.
\label{IPR}
\end{eqnarray}
Where $c_{n\vert 0}(t)$ are the time dependent correlation amplitudes of the one-body wave function in the local (computational) basis. Furthermore, our numerical simulations show that the variance of the \( Q_n\) distribution is identical to the variance of the excitation. That is,

\begin{equation}
\sum_n Q_n(t)n^2=\sum_n S_n(t)n^2-(\sum_n S_n(t)n)^2.
\label{varianzaS}
\end{equation}It is important to note that we are not calculating the usual inverse participation ratio of the eigenstates but that of the actual polarization excitation as it evolves. Our magnitude represents how well the initial \textbf{excitation} spreads over the chain. While for $U=0$, $\tilde Q_0$ coincides with the $\rm{IPR}_t$, this is not exactly true in presence of interactions ($U\neq0$). There, our magnitude still represents how much the magnetization spreads in real space but the appearance of higher terms in the $Q_n$ spectrum is indicative of some growth of the Hilbert space which may not correspond with a polarization dynamics. 

The short-time growth of OTO commutators have been proposed and used as a measure of scrambling and onset of quantum chaos. However, it has been the details of the long time behavior of different OTOCs which captured wide attention as tools to study chaos, thermalization, and localization. \cite{HGUR19,RUR18,Gm+Wi18,Le+Re19}. These can show the variety of notable interference phenomena that characterize correlations functions, such as quantum beats \cite{FGmJaWi19}, survival collapse \cite{RfPa06} at intermediate times, notable correlation holes\cite{TS17} at long times and mesoscopic echoes at the Heisenberg time \cite{PLU95,ADLPa10}.
 
In the next sections we present the numerical study of the dynamics of excitations in the \textbf{ interacting Harper-Hofstadter-Aubry-André spin model } using as main quantifiers the observable that result from the ZOGE sequence. For $U=0$ the simulations were made by exact diagonalization while for $U\neq0$ a Trotter-Suzuki dynamics\cite{DRae83,DBZgPa13} is employed with the quantum parallelism method to evaluate self-averaging quantum observables\cite{ADLP08}. This last drastically reduces the calculation time with respect to the standard density matrix approaches.

\begin{figure}[t]
\centering
\includegraphics[width=0.48\textwidth]{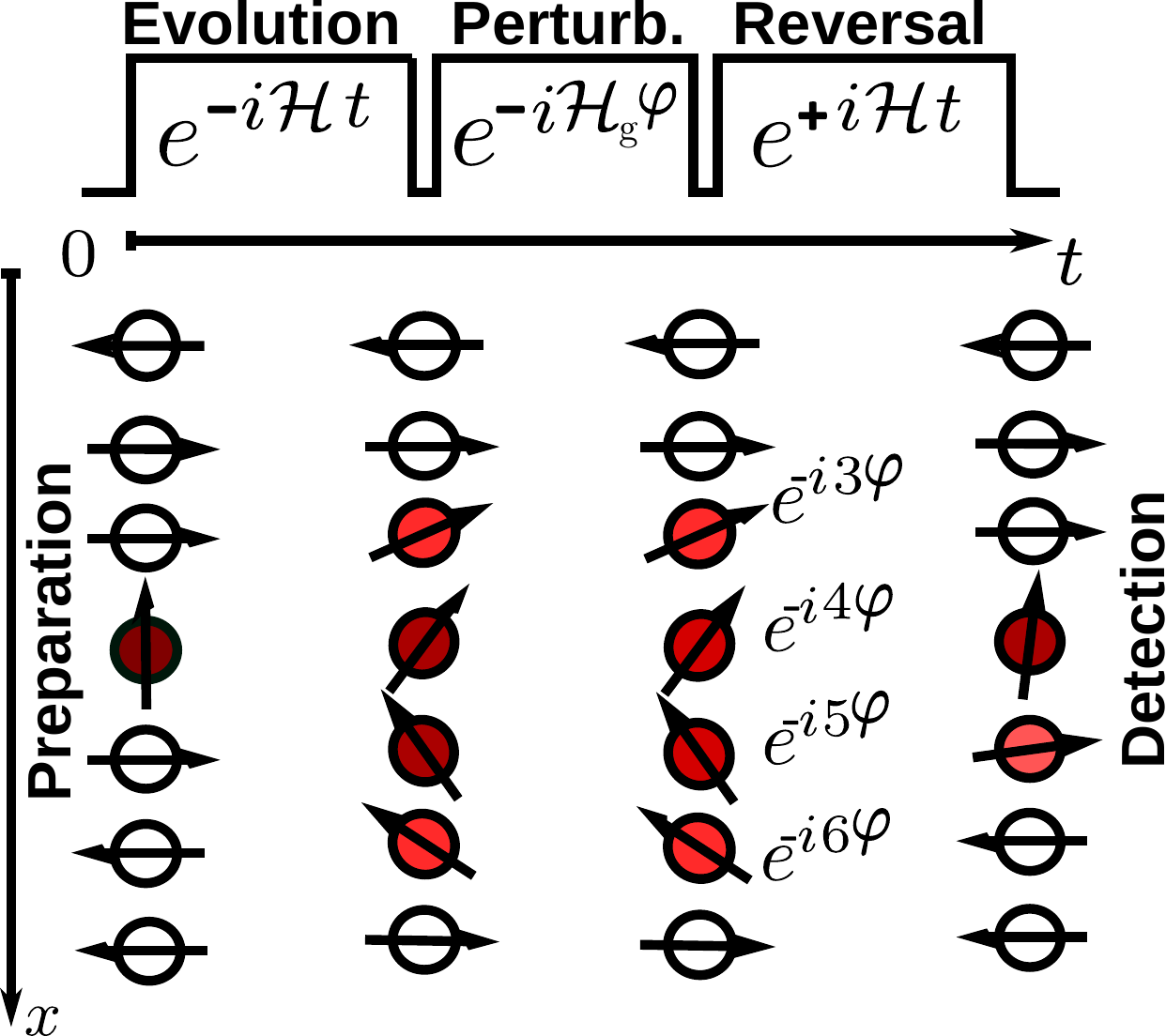}
\caption{ Schematic representation of the state of spin chain at different times. It describes the sequence of Hamiltonians that operate on an initially localized excitation at the central spin. The phases $\varphi$ result from different magnetic field gradient pulses. Finally, a local measurement is implemented at the same spin. A Fourier transform of these observables provides the \textit{zero order gradient entanglement} $\tilde Q_0$ (ZOGE) which, in a non-interacting Fermionic representation, is an inverse participation ratio.}
\label{Modelo}
\end{figure}

\section{The dynamics of a local excitation in a one-body framework.}

We pointed above that in absence of the Ising term, the size of the Hilbert space has the same dimension as the maximally negative spin projection Hilbert subspace of the interacting system\cite{Ma+Er97,PLU95}. This allows us to explore our proposal without too much computational cost in a system where the phase transition is theoretically well defined and can be observed numerically with good precision in long chains. 

\begin{figure}[t]
\centering
\includegraphics[width=0.48\textwidth]{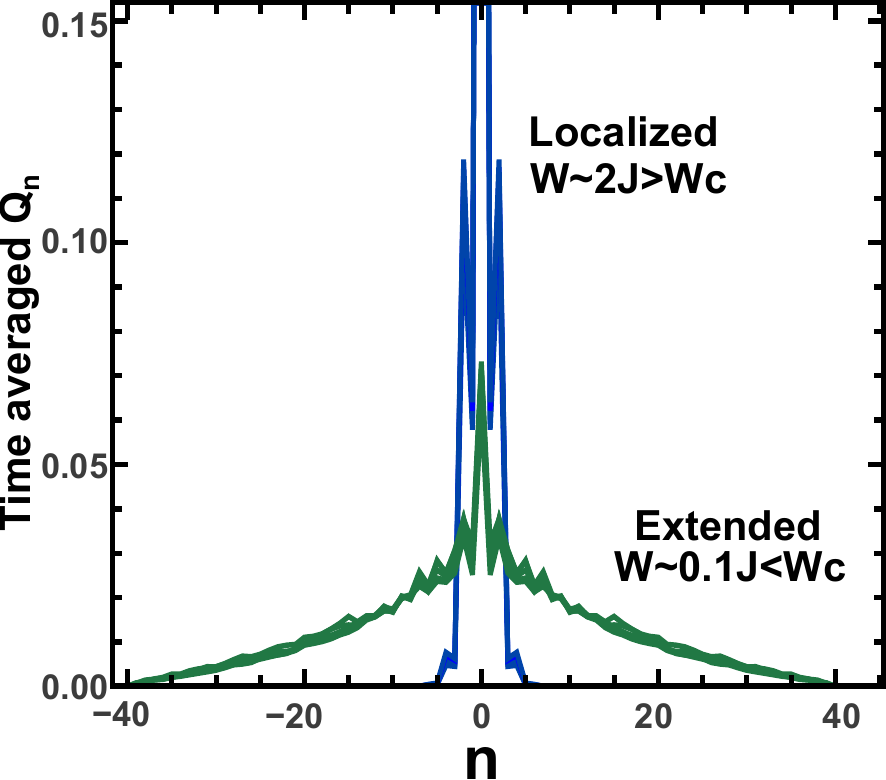}
\caption{Time-averaged Fourier spectrum $\tilde Q_n$ of gradient entanglement amplitudes (GEA) showing that in the extended phase (green) it has a wider spreading that in the localized phase (blue).}
\label{spectrum}
\end{figure}

In Fig. \ref{Modelo} we show a scheme of the system at different times. The initial state is prepared as an imposed local magnetization. The many-body dynamics leads to the spreading of this excitation. As it is shown, the different local components of the magnetization acquire a phase which is proportional to the distance to the original site of the excitation. As a consequence of the phase shifts, some magnetization fails to come back to the initial site in spite of time reversal. By repeating this process for several phases $\varphi$ between $0$ and $2\pi$, one can extract the amplitudes $\tilde Q_n$. If one uses longer chain lengths, $N$, one also would need smaller phase increments to reach gradient entanglement amplitudes ($\tilde Q_M$) that correspond to longer distances ($M$). Thus, more phases $\varphi$ are needed to extract the complete spectrum. These amplitudes, which encode how far the magnetization has gone, reflect the localized or extended state of the system as it is shown in Fig. \ref{spectrum}. There, we observe how the time averaged spectrum is wider in the extended phase.

\subsubsection{Initial condition}

\begin{figure}[t]
\centering
\includegraphics[scale=0.9]{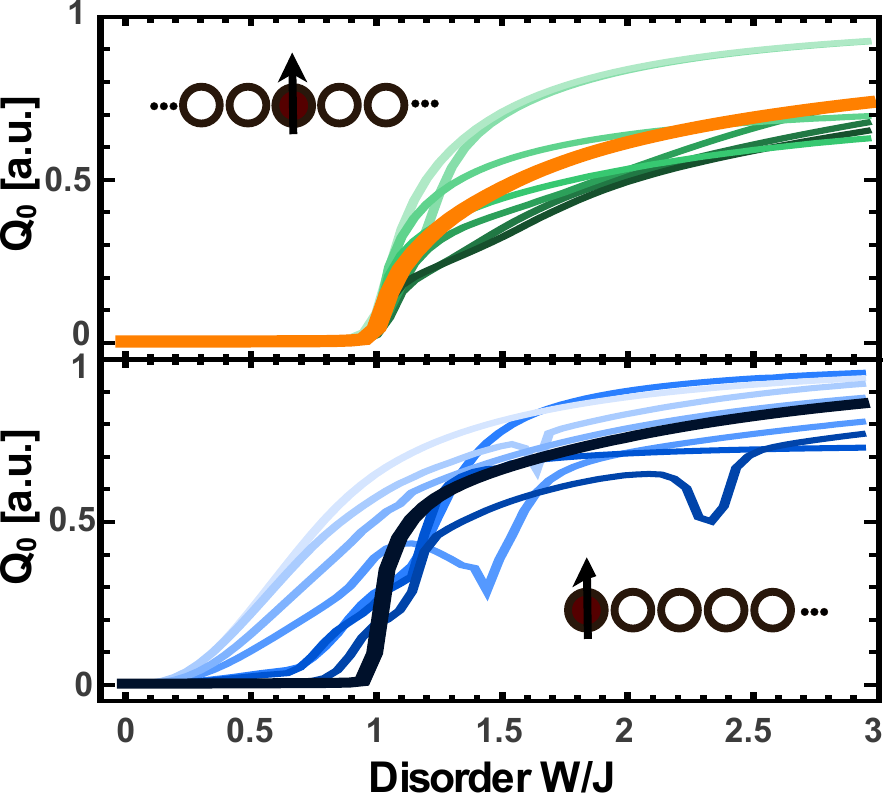}
\caption{Equilibrium value of $Q_0$ as a function of $W$ for different disorder realizations ($\phi$) in a 500 spin chain. Above (Green): The initial excitation is placed in the middle of the chain. Bottom (Blue): The excitation is placed in a extreme of the chain.}
\label{CI500}
\end{figure}

In the ZOGE sequence, one of the free variables is the location of the initial condition. We have simulated the evolution of $Q_0$ in two conditions. One corresponds to a local excitation in the middle of the chain and the other when it is in an extreme of the chain. Fig. \ref{CI500} shows the equilibrium value of this magnitude (averaged over time) as a function of the modulation strength ($W$) for different realizations ($\phi$). We observe that when the initial condition is in the middle of the chain, for $W<W_c$ the magnetization can always spread all over the chain, and for $W>W_c$ the magnetization remains near the initial excitation. This is consistent with the Aubry-André transition. However, if the excitation is placed at an extreme of the chain we observe that even for modulations below the critical one, there are systems whose local magnetization can not spread over the whole chain. This effect is shown in Fig. \ref{CI500} where only one disorder realization (black,$\phi=0$) reflects the expected critical behavior at $W_c=1$. In the rest of the realizations, the excitation remains localized for $W<W_c$.

\begin{figure}[t]
\centering
\includegraphics[scale=0.9]{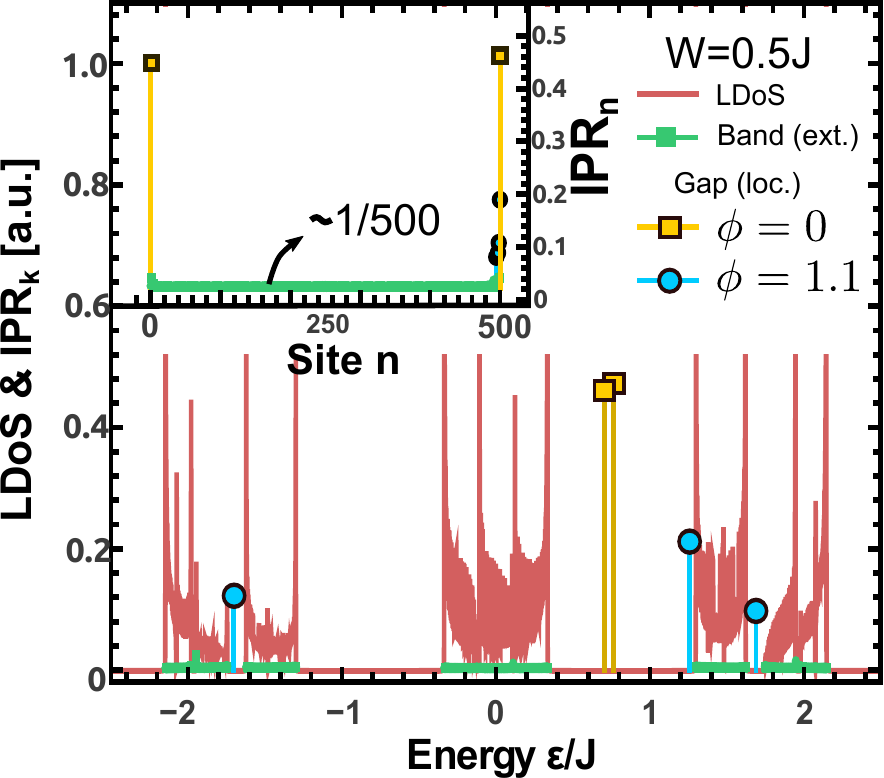}
\caption{Main: Local density of states of the Hamiltonian $\mathcal{H}$ with $W=0.5<W_c$ (Red). $\mathrm{IPR}_k$ of the eigenvectors of $\mathcal{H}$ with $W=0.5J<W_c$ and $N=500$ for two disorder realizations. The nonlocalized states are shown in green (bulk states) while the localized states of each realization are shown in light blue ($\phi=0$) and orange ($\phi=7\pi/20$). Inset: Participation ratio of site states over the eigenstates $\mathrm{IPR}_n$ (same color scheme).}
\label{ONEBODYD}
\end{figure}

To understand this effect we have studied the eigenenergies and eigenvectors of the Hamiltonian in a 500 spin chain. Fig. \ref{ONEBODYD} shows the local density of states of a long chain (calculated using the decimation method \cite{PWA83,PM01}) and the degree of delocalization of the eigenvectors as a function of the energy. This last is evaluated through the quantities:
$\mathrm{IPR}_k =\sum_{n} |a_{kn}|^4$ and $\mathrm{IPR}_n=\sum_{k} |a_{kn}|^4$.

When the chain is finite (500 spins) some eigenstates have a higher $\mathrm{IPR}_k$, which means there are localized eigenstates. These edge states appear in the gaped region of an infinite chain and are a finite size effect. Furthermore, if we calculate the participation ratio of site states over the eigenstates we observe that these localized states are in the extremes of the chain. Thus, they should be distinguished from highly improbable mini-bands whose main weight lies at sites where the on-site potential is anti-symmetric \cite{PWA83}. These, rare low mobility solitons may show up in the bulk and are not expected to contribute with an appreciable effect on excitation dynamics. It is interesting to note that the phase $\phi=0$ has very special properties. In that case, there is no localized state at such edge, but there are localized states supported at the opposite extreme instead. Alternatively, for arbitrary phases, say $\phi=7\pi/20$, both extremes support edge states. The absence of finite size effects for $\phi=0$ ensures that this dynamical IPR will show localization only for $W>W_c$. However, other initial phases, such as $\phi=7\pi/20$, yield edge states for $W<W_c$. This uncontrolled variability motivated us to use initial conditions in the middle of the chain as a safe representative of the phase transition in the thermodynamic limit.

\subsubsection{Scaling}

\begin{figure}[t]
\centering
\includegraphics[scale=0.9]{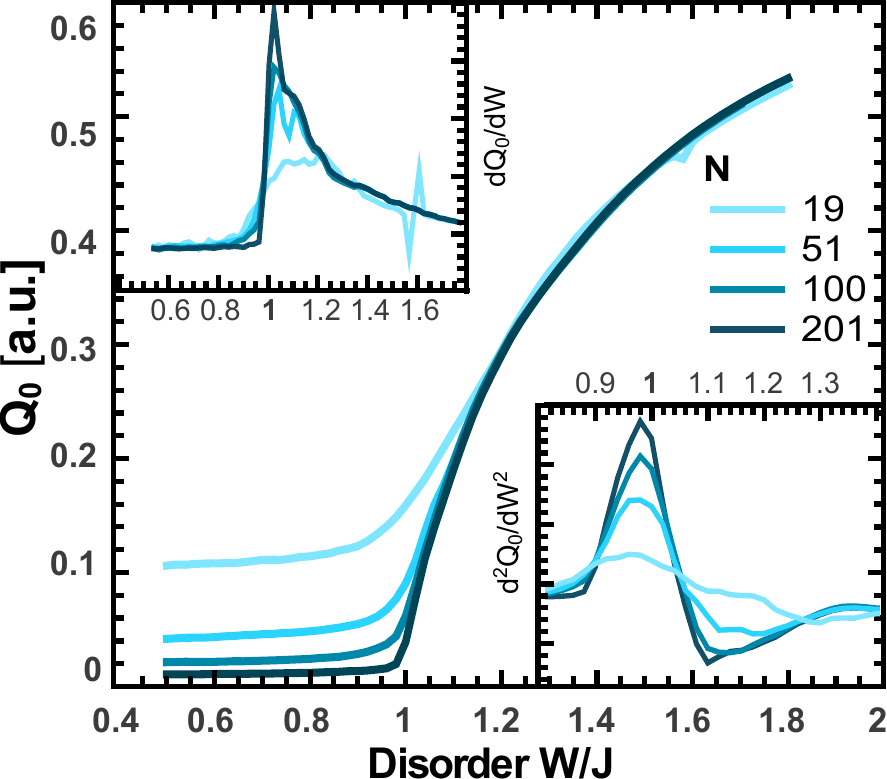}
\caption{Equilibrium value of $Q_0$ averaged over 10 disorder realizations ($\phi$) as a function of $W$ for $N={19,51,100,201}$. Insets: First and second derivative $\dfrac{dQ_0}{dW}$, $\dfrac{d^2Q_0}{dW^2}$, which are used to enclose the critical value $W_c$.}
\label{ONEBODYN}
\end{figure}

A detailed study of a phase transition usually requires simulations with different systems sizes that make possible a finite size scaling analysis. However, in the interactive system, our maximum $N$ achievable is restricted to about 23 spins making these strategies not accessible. Still we would like to find bounds for the critical parameters. In order to develop such procedure we first restrict ourselves to the non-interacting case. Fig. \ref{ONEBODYN} shows the equilibrium value of $Q_0$ averaged over 10 disorder realizations ($\phi$) as a function of $W$ for several chain lengths ($N$). We observe that in the extended regime ($W\ll 1$) the asymptotic $Q_0$ scales as $1/N$. This behavior is consistent with the more standard definition of IPR in a delocalize regime \cite{McKr93,DoCa19}. It shows that the initial excitation can spread all over the chain. On the contrary, in the localized phase there is only a weak dependence on $N$. This is because the magnetization can only spread within a localization length.

\begin{figure}[t]
\centering
\includegraphics[scale=0.9]{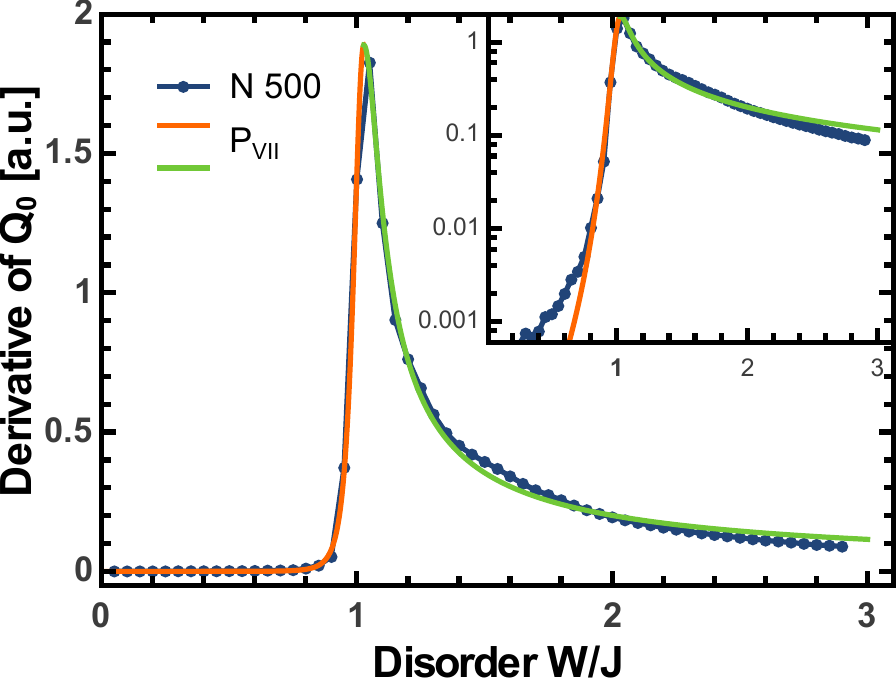}
\caption{Derivative $\dfrac{dQ_0}{dW}$ and split fitting. The asymmetry of the fitted function is remarked in orange (left) and green (right).}
\label{ONEBODYFIT}
\end{figure}

We first study the numerical first derivatives ($\dfrac{dQ_0}{dW}$), which are shown in the upper-left inset of Fig. \ref{ONEBODYN}. Their maximum is at the modulation strength where the original curve changes its concavity. One observes that as $N$ increases, these maxima always yield an upper bound to the $W_c$. An analysis of the second derivatives (Fig. \ref{ONEBODYN} lower-right inset, $\dfrac{dQ^2_0}{dW^2}$) shows a maximum that, for all the system sizes, provide a lower bound to the critical value. Instead of working with raw numerical data, one get less noisy results if we extract these maxima by fitting a split Pearson type VII function:
$$P_{\rm{VII}}(x;\vec a)= \left\{ \begin{array}{c}
  \frac{a_0}{[1+(\frac{x-a_1}{a_2})^2(2^{1/a_3}-1)]^{a_3}} x<a_1 \\
  \\
\frac{a_0}{[1+(\frac{x-a_1}{a_4})^2(2^{1/a_5}-1)]^{a_5}} x>a_1.
\end{array} \right.
$$
 An example of this fitting is shown in Fig. \ref{ONEBODYFIT} in a system of length $N=500$. This procedure allows us to handle the different behavior at both sides of the transition for both the non-interacting and the interacting cases.

\section{Local excitation dynamics under many-body interactions.}

\begin{figure}[t]
\centering
\includegraphics[scale=0.9]{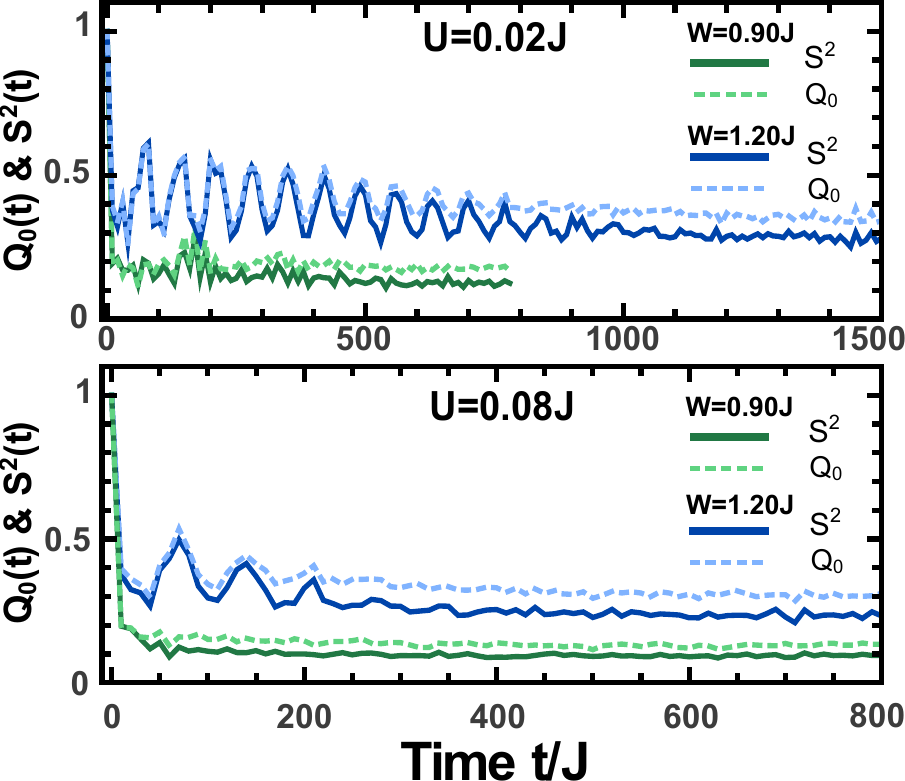}
\caption{Numerical simulation of the \textit{zeroth-order gradient entanglement} ($Q_0$, solid lines) and the sum of the square magnetization ($S^2$, pointed lines) for two interaction strength $U=0.02J$ (top) and $U=0.08J$ (bottom). The colour of the curves describe the disorder strength, $W=0.90J$ (green) and $W=1.20J$ (blue), \textit{i.e.} extended and localized respectively. The simulations correspond to a 13-spin-chain with an initial excitation in the 7th spin.}
\label{1}
\end{figure}
 When the interaction is turned on, in Trotter approximation the initial excitation $S_0^z(0)$ starts its spreading \textbf{through the Liouville space} as $\hat S_0^z(\delta t)= \hat S_0^z(0) \cos(\delta t/T_2)-\rm{i}[\mathcal{\hat H}, \hat S_0^z(0)]\sin(\delta t/T_2)$. Although $[ \mathcal{\hat H},\hat S^z]=0$, \textit{i.e.} the sub-spaces with different total magnetization $M$ do not mix among themselves, at later times the terms \([\mathcal{\hat H}, \hat S_n^z(t)]\) produce multi-spin terms within each subspace. These terms, add positive contributions to $\tilde Q_0$, which lead to a difference with the IPR and its many-body equivalent, the sum of the squared magnetizations, 
 \begin{equation}
S^2(t) =\sum_{n} \langle \hat S_{n}^z(t)\rangle_{\beta}^2.
\end{equation} 
Through numerical simulations for various chain sizes, we observed that this small difference between \(Q_0(t)\) and \(S^2(t)\) does not affect the localization/delocalization analysis. Though some monotonic differences appear, $Q_0$ follows the dynamical behavior of the $S^2$. Fig. \ref{1} shows the amplitude of the ZOGE ($\tilde Q_0$, solid lines) and the sum of the squared magnetizations ($S^2$, points) for two modulation strengths, $W=0.90J$ and $W=1.20J$, which correspond to the extended and localized regimes respectively, and interactions strengths enough to cause some delocalization $U=0.02J$ and $U=0.08J$. We have observed that, in the more localized cases, these quantities need time to make their difference more evident. This long time scale allows the interaction to smear out both, the breathing modes of the localized phase (oscillations in the blue curves) and mesoscopic echoes of the extended phase (revivals in the green curves).\cite{PLU95,PUL96} In particular, when the system is in the localized phase $S^2(t)$ and $\tilde Q_0(t)$ coincide for longer times.
 
These results (Fig. \ref{1}) were obtained with an initial state that belongs to the subspace of total magnetization $M=1/2$. This is equivalent to a system with $(N+1)/2$ Fermions, in a chain with an odd number of sites $N$. Thus, one optimizes the computational resources while working in the most interacting subspace. This subspace is representative of an NMR experiment \cite{LUPa98} and it is also typically used in the cold atoms experiments as well as other numerical studies\cite{Xu+SwSa19,MR15,Sc+Bl15,Ag+Dm15,Lu+Bl17}. We also confirmed that $\tilde Q_0$ and $S^2$ follow identical dynamics for the subspace with total spin projection $M=1-N/2$, where the excitation dynamics becomes one-body. They start to differ as the subspace index decreases. This means that the discrepancy observed in Fig. \ref{1} is the greatest that could be observed in this system.

The main experimental utility of our sequence is that it allows us to obtain information about the scrambling on the whole system just by probing a few individual spins of the ensemble. This is an almost unique possibility given by the use of a rare nucleus as a local probe in conjunction with the Loschmidt Echo/OTOC procedure. The rare nucleus first injects and, later on, probes the magnetization in its directly connected nucleus\cite{LUPa98}. From this initial nucleus, the LE/OTOC sequence allows the dynamical exploration of the rest of the system. This feature circumvents the necessity of multiple single measurements of the local components of the magnetization, a quite exceptional achievement of few NMR experiments\cite{Ma+Er97,A+LPa10}. However, the numerical simulations can be done at a lower cost without implementing time-reversal, by evaluating just the sum of the squared magnetization under a forward dynamics.

The evolution of $S^2$ was implemented in a 19-spin chain for several values of the parameters $U$ and $W$, with an initial excitation placed at the center of the chain which, as discussed, allows us to avoid the edge effects. As it is pointed out in Ref. \cite{Xu+SwSa19}, the HHAA model with interaction might not maintain the property that the extended-localized transition occurs simultaneously for all the single-particle eigen-energies. That is because, in a mean field sense, a renormalized single-particle potential would not maintain the self-duality \cite{WP84,LCW84}.

 This would mean that for certain parameter ranges both localized and delocalized single particle states could coexist at different energies. Still, our numerical data maintains the advantage of the HHAA model of precluding the need to perform extensive ensemble averages to observe the phase transition. Our simulations show that this is generally true. An exception occurs when the excitation starts at an extreme of the chain, where the edge states effects can be misinterpreted as bulk localization. Nevertheless, in the localized phase, even with initial conditions in the middle of the chain, the elimination of local fluctuations at long times requires ensemble averages as each particular realization has different details of the $S^2$ dynamics.

\begin{figure}[t]
\centering
\includegraphics[scale=1]{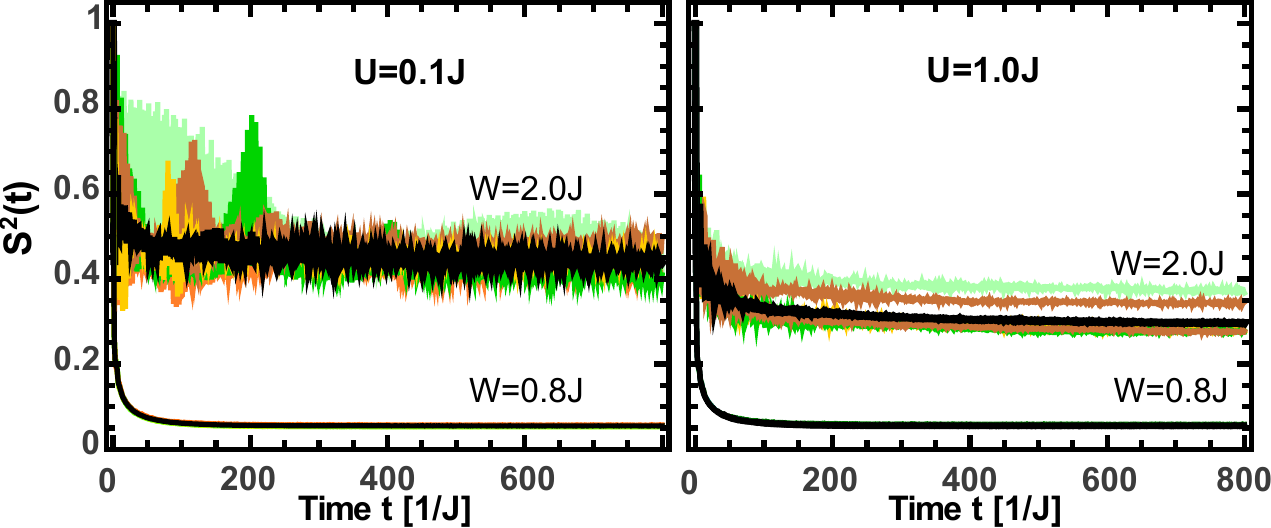}
\caption{Time evolution of sum of square magnetization ($S^2$) in the extended and localized phases for five disorder realizations (colour on-line) and the ensemble average (black curve). Left panel: Interaction strength $U=0.1J$. Right panel: Interaction strength $U=1.0J$}
\label{a}
\end{figure}

Fig. \ref{a} shows the time evolution of the sum of the squared magnetizations $S^2$ in the extended and localized phases for two interaction strengths and various disorder realizations. The effects discussed in the former paragraph can be observed in Fig. \ref{CI500} (the one-body equilibrium values) and also in Fig. \ref{a}. We observe that in the extended phase ($W=0.8J$) the differences between the individual dynamics and equilibrium value obtained from different disorder realizations becomes negligible. In the localized phase this difference may be stronger. In particular, when the interaction strength $U$ is small, the recurrences in \(S^2(t)\) make necessary to obtain the equilibrium value by performing extensive disorder averages (black lines). Once these oscillations have been averaged out the asymptotic values and the decay law manifest more clearly. 

\subsubsection{The asymptotic polarization.}

\begin{figure}[t]
\centering
\includegraphics[scale=0.9]{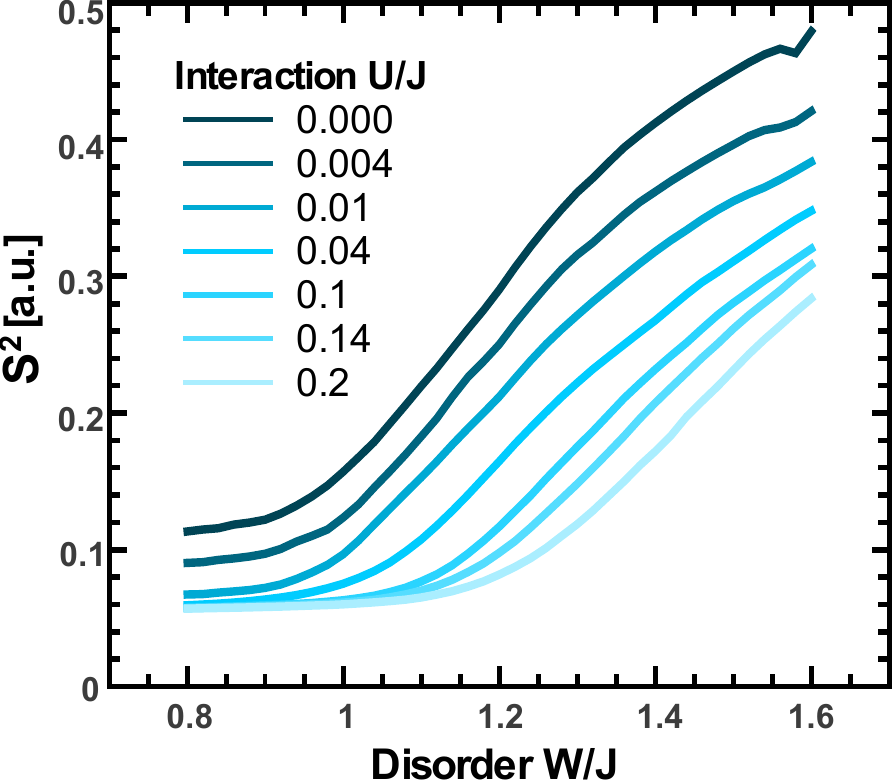}
\caption{Equilibrium value of the disorder averaged sum of square magnetization ($ S^2 =\langle \sum_{j} \langle S_{j}^z\rangle_\beta^2\rangle_\phi $) as a function of the disorder strength ($W$) for several interactions strength from $U=0J$ to $U=0.2J$ in a $N=19$ chain. }
\label{2}
\end{figure}

In Fig. \ref{2} we display the (disorder averaged) asymptotic values $ S^2 $ as a function of $W$ for several interaction strengths. For $U=0$ we have observed that this quantity increases rapidly when $W$ surpass $W_c=J$ indicating the phase transition between extended and localized states.

As the interaction strength grows, two effects appear. First, for both the extended and the localized regimes, the value of $S^2$ decreases very abruptly with the interaction. This indicates that even small interactions are very effective in assisting the redistribution of the polarization preventing interferences and recurrences as the polarization echoes. Something similar is observed when the precise ballistic propagation of wave packets becomes destroyed by small values of $W$. Secondly, the growth of $S^2$, which is identified as a signature of localization, occurs for higher disorder strength as the interaction increase. This is indicative of the delocalizing effect of many-body interactions. 

In order to have a first glimpse on the influence of \(U \) in the critical value of $W$ we have analyzed the derivative $d( S^2)/dW$ and fitted the center of the observed peaks and also the maximum of the second derivative $d^2(S^2)/dW^2$. As we pointed for the non-interacting case, the maximum of the first derivative can be interpreted, even for our small \(N\), as an upper bound for $W_c$. Conversely, the second derivative maximum provides a lower bound for $W_c$. Notably, as the interactions are turned-on U the critical value of $W$ increases very little, however as \(U > 0.04J\) the critical value tends to grow more rapidly. This is in accordance with the experimental observation \cite{Sc+Bl15}. There are previous numerical results that include an extra degree of spin and second neighbor interaction which should represent better the experimental system. However, they do not have the same precision at assessing the weak interaction range \cite{MR15}. Besides, in this regime, the difference between models is deemed to be less important.

\begin{figure}[t]
\centering
\includegraphics[scale=0.9]{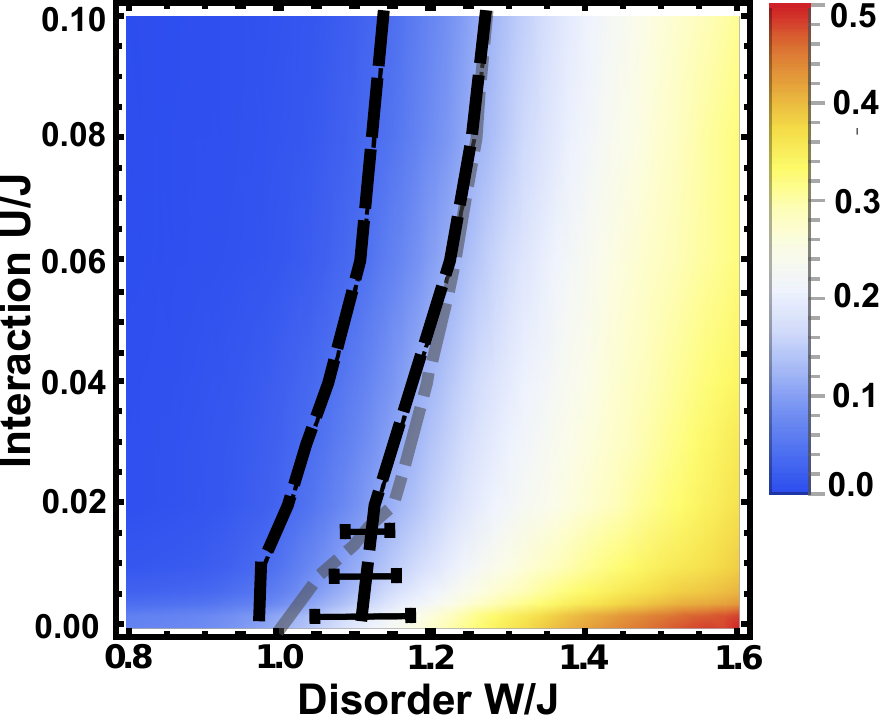}
\caption{Density map of $ S^2=\sum_{j} \langle \langle S_{j}^z\rangle_\beta^2\rangle_\phi$ versus $U$ and $W$. "Schematic phase
diagram for the system. In the ergodic-delocalized phase (blue), the initial $S^2$ quickly decays, whereas it persists
for long times in the nonergodic, localized phase (yellow). The black-dashed lines show the bounds founded for the critical value by analyzing the first and second derivatives of $S^2$. The grey line corresponds to the contour line beginning in the critical value for $U=0$.}
\label{10}
\end{figure}

The effects and bounds described above can be better understood in the color contour plot of \(S^2\) on the parameter space, \((W,U)\), around the Aubry-Andre transition (Fig. \ref{10}). Here, the blue parts of the diagram represent small values of $S^2$, that is a delocalized/ergodic phase while the yellow-red parts of the plot correspond to the localized/non-ergodic regime. The upper and lower bounds are shown as black dashed lines, while the grey-dashed line is contour-line starting at the non-interacting critical value (W=1,U=0). Thus, it is clear how small interactions move the localized phase to higher values of $W$. 

\subsubsection{The survival probability and the OTOC scrambling dynamics.}

\begin{figure}[t]
\centering
\includegraphics[scale=0.9]{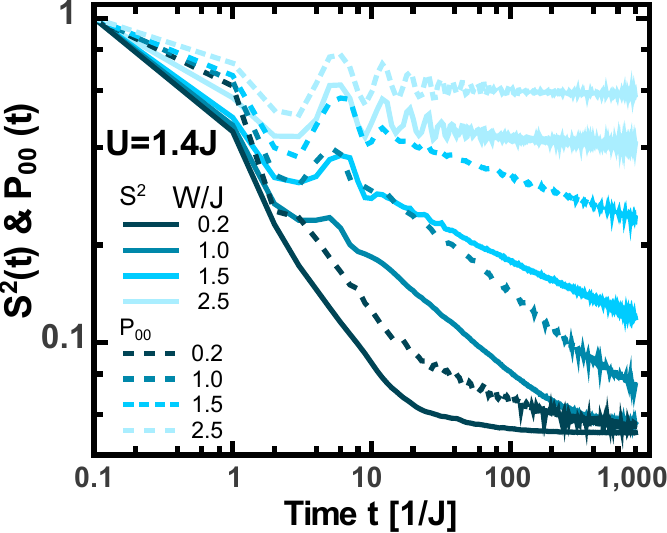}
\caption{Time evolution of $S^2(t)$ and $P_{00}(t)$ (solid and dashed lines) for $U=1.4J$ and several disorder realizations. This log-log scale captures the power decay for the intermediate time regime, but misses (due to lack of precision) the very short time decay, \(\propto 1-\frac{1}{2}(Jt)^2\), that characterizes short times.}
\label{7b}
\end{figure}
\begin{figure}[t]
\centering
\includegraphics[scale=1]{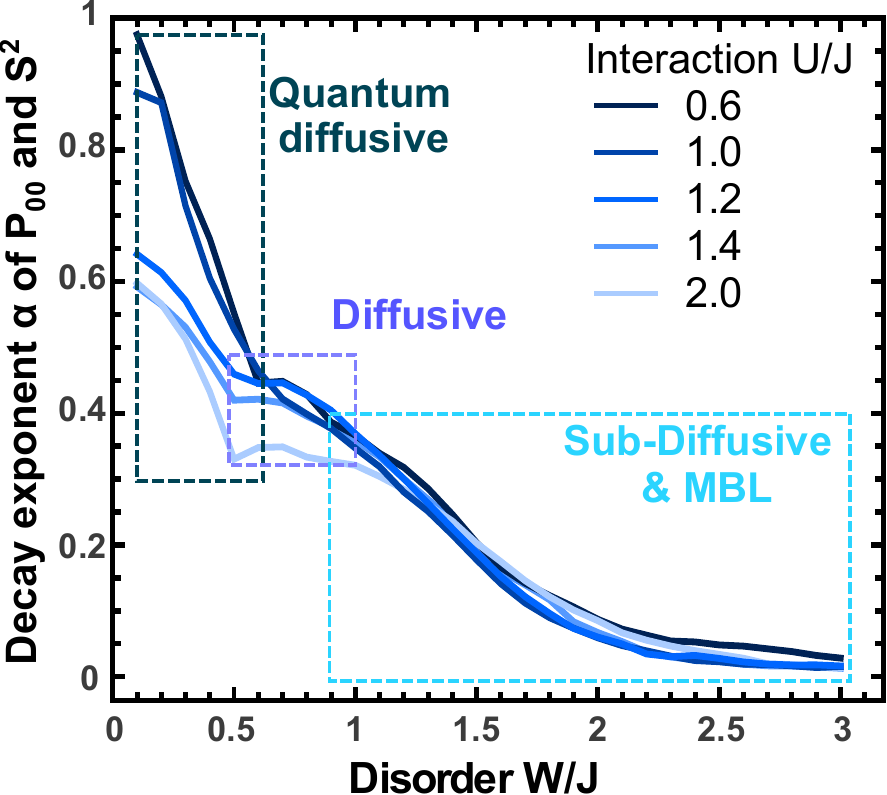}
\caption{Exponent $\alpha$ of the power law decay ($1/t^\alpha$) of a local excitation as described by \(P_{00}(t)\) and \(S^2(t)\). The  $\alpha$'s are shown as a function of the disorder $W$ for different interaction strength $U$. For ideal ballistic quantum diffusion the exponent is \(\alpha=1\), scattering processes around the critical disorder lead to a diffusion \(\alpha=1/2\)  regime that is quite robust under many-body interactions. In our finite system, many-body localization is seen as sub-diffusion, an exponent that banishes for strong disorder. }
\label{alpha}
\end{figure}

The dynamical behavior of different initial conditions in the presence of interactions and disorder have been widely studied, showing a variety of results depending on the parameters. In particular, a slow (sub-diffusive) dynamics has been observed before the appearance of many-body localization phase. \cite{Ag+Dm15,LKKRK17,Lu+Bl17} The explanation of this dynamics is still a subject of discussion. For random disorder, theoretical studies have predicted a Griffiths regime on the thermal side of the transition \cite{Gri69}, where the dynamics is dominated by rare spatial regions with anomalously large escape times. However, this picture may not apply when there are long-range correlations in the underlying disorder potential.\cite{G+HuM16,CLO15} 

The regime of slow excitation dynamics also appears in our simulations of $S^2(t)$. In Fig. \ref{7b}, we show this sub-diffusive decay for an interaction strength $U=1.4J$ and several disorder amplitudes. The solid lines represent the studied $S^2(t)$, while the dotted lines show the survival probability $P_{00}(t)$. The power law decay results from the prevalence of the return amplitude, as carefully described in Ref. \cite{RfPa06}. We observe that while the initial decay of $S^2(t)$ is faster than the decay of $P_{00}(t)$, once the sub-diffusive regime is reached both decays can be approximately described by the same exponent. This can be understood within simplifying model. Let's suppose a standard diffusion given by $P_{x0}(t)=1/({2\pi \sigma^2})^{1/2}e^{-(x/\sqrt{2}\sigma)^2}$, where $\sigma \propto [\sqrt{2}Dt]^{\alpha}$. A case of \(\alpha=1\) would describe a quasi-ballistic quantum diffusion. Once the scattering processes become relevant, the exponent should stabilize at around \(\alpha =1/2\), that is indicative of diffusion in 1\(d\). In our case the Gaussian law represents the conserved magnetization, and \(\alpha =d^*/2\) could represent a diffusion in a restricted fractal space of dimension \(d^*\leq 1\)\cite{CuPa00}. In this simple model, the survival probability is given by $P_{00}(t)=1/\sqrt{2\pi \sigma^2} \propto t^{-\alpha}$. The magnitude $S^2(t)$ in this model would be the integral of the squared magnetization $\int_{-\infty}^{\infty}[P_{x0}(t)]^2\mathrm{d} x=\dfrac{1}{2\sqrt{\pi\sigma^2}}\propto t^{-\alpha}$. As the localization conspires against diffusion, \(\alpha\) becomes smaller than \(1/2\) accounting for a progressively sub-diffusive dynamics. 

The validity of the previous picture is reinforced by the agreement of the exponents that characterize our dynamical observables: the local survival probability and the \textit{degree of scrambling} as quantified by the integral of the squared magnetization. The observed behavior fully agrees with the physical processes we expect in the various regimes of the system as we change the ``disorder" strength. For finite but weak disorder strength, the excitation spreading slows down from quantum diffusion to classical diffusion (Fig. \ref{alpha}). There, a diffusive dynamics is stable for a small range disorders while the spreading can not reach the whole 1\(d\) system but a fraction of it characterized by a fractal dimension \(d^*\leq 1\). Further increase of disorder continuously slows down the scrambling until it enters in the MBL phase. There is another interesting piece of information that can be drawn from the asymptotic data of the diffusive and sub-diffusive regimes as we also get \(S^2\simeq L^{-d^*}\). Indeed, this turns out to be fully consistent with the exponent of the decay of the survival probability and the squared magnetization integral. By considering times between $t\sim10/J$ and $t\sim 500/J$, we were able to fit the power law decays of the return probability and the sum of the squared magnetization. We obtained the same exponent $\alpha$ of both cases, reinforcing the above observation. Indeed, both exponents have the same dependence on the disorder strength with a small difference in the transition region. Their decrease with increasing disorder indicates a progressively slower equilibration (Fig. \ref{alpha}).

\section{Conclusions}

We proposed and discussed a sequence to measure the asymptotic delocalization of a local excitation through the evaluation of $\tilde Q_0$, an OTOC that we denominated \textit{zeroth-order gradient entanglement} or ZOGE. It provides a practical method to analyze the localization/delocalization transition and, more generally, a tool to quantify the excitation dynamics as it becomes non-ergodic. The only requirement is that one needs access to a local observable and that the global dynamics could be time-reversed, both of which are quite standard in NMR and, for extension, in other experimental techniques. The most appealing property is that, in the limit of negligible interaction, $\tilde Q_0$ becomes precisely an inverse participation ratio, IPR, of the polarization and also it approximates its generalized definition $S^2$ for the interacting case. While this observable does not satisfy a precise conservation law, it is somehow equivalent to the integral of a square charge density in a Fermionic gas. This correspondence also provides a sound physical interpretation for this particular OTOC. 

The similarity between $\tilde Q_0$ and $S^2$ allows us to estimate the long-time value of $\tilde Q_0$ through the simpler calculation of $S^2$ with much less computational cost. We have made an analysis of $S^2$ in the surrounding of the critical point $W_c=J$ for small interaction strength ($U<<J$), which allows us to track the critical disorder as $U$ increases. We observe how the critical disorder increases with $U$, indicating that many-body interactions, acting as a decoherence source, diminish the interferences underlying the localization phenomenon. Interactions, then, assist in the scrambling of the magnetization within the fraction of the space available by the disorder.

In our numerical strategy, we study the asymptotic behavior of the excitations and our results are able to discriminate between localized and extended states regardless of the existence of slow decays. Although we observe parametric regions with slow processes in agreement to the existence of a slow \textit{S phase} reported in \cite{Xu+SwSa19}, in our small systems we were not compelled to assign a different phase. 

We also notice that there is much room for further exploration on the information that can be extracted from the whole \textit{high order gradient entanglement spectrum}. However, in this work, the relevant information was inferred from the numerical solutions of the asymptotic value of $S^2$, which has with much lower computational cost. In an actual experiment, one can take advantage not only of $\tilde Q_0$ but of all the $\tilde Q_n$ for different \(n\)'s. Indeed, we have tested the usefulness of the variance of the spectrum for $U=0$. Such magnitude coincides with the spatial variance of the excitation as measured by Eq. \ref{varianzaS}, thus, we expect that this property holds, as an approximation, in the interacting case. In this regime, the increasing differences in the spectrum, like the appearance of higher values of $\tilde Q_n$ with respect to the non-interacting case, could find further use to decode information about the actual occupation of the Hilbert space. An important property of \(S^2\) and, through their correspondence, of $\tilde Q_0$ is that it measures how a real space observable spreads. Accordingly, it does not depend directly on the size of the Hilbert space. 
Since polarization is a conserved magnitude in a many-body spin system, one can use it to define a Renyi entropy\cite{UPaL98}. The connection of this last with an inverse participation ratio (IPR), obtained as the sum of the square of local polarizations, makes it a natural magnitude to quantify the Many-Body-Localization in a chain of interacting spins or related models. However, until now the experimental implementations have been limited to small molecular systems \cite{LiZeDu17} or indirect observables that are only a proxy for the localization properties of disordered spin chains \cite{WeiChCa18}. Our work sought to overcome these limitations by proposing a strategy that focuses on the simple IPR/Renyi entropy of the non-interacting case with a procedure that conserves its potential in the MBL regime.

Finally, we have found that the long-term dynamical behavior of $S^2$ and $P_{00}$ is driven by the same decay exponents. This relation, which we also derived from a simple toy model, shows how the dynamics change from quantum diffusive to sub-diffusive and then to many-body localization as the disorder increases. This slow decay is consistent with previous results \cite{Ag+Dm15,LKKRK17,Lu+Bl17,Xu+SwSa19}. Furthermore, the power law exponent $\alpha$ in the sub-diffusive regime can be directly connected with a non-ergodic dynamics restricted to a portion of the system with fractal dimension $d^{*}<1$\cite{CuPa00}.

\section{Acknowledgements}
This work has received financial support through grants and fellowships from Argentine agencies: SeCyT-UNC, CONICET and ANPCyT (PICT 2017-2467).

\bibliographystyle{elsarticle-num-names}
\bibliography{main.bib}

\end{document}